\documentclass[lineno]{biometrika}
\usepackage{graphicx} 
\usepackage{multirow}
\makeindex
\usepackage{amsmath}

\usepackage{times}
\usepackage{bm}
\usepackage{natbib}
\usepackage[margin=0.5 cm]{caption}

\usepackage[plain,noend]{algorithm2e}

\makeatletter
\renewcommand{\algocf@captiontext}[2]{#1\algocf@typo. \AlCapFnt{}#2} 
\def\@algocf@capt@plain{top}
\renewcommand{\algocf@makecaption}[2]{%
  \addtolength{\hsize}{\algomargin}%
  \sbox\@tempboxa{\algocf@captiontext{#1}{#2}}%
  \ifdim\wd\@tempboxa >\hsize
    \hskip .5\algomargin%
    \parbox[t]{\hsize}{\algocf@captiontext{#1}{#2}}
  \else%
    \global\@minipagefalse%
    \hbox to\hsize{\box\@tempboxa}
  \fi%
  \addtolength{\hsize}{-\algomargin}%
}
\makeatother

\begin{document}

\jyear{2012}
\jvol{99}
\jnum{1}
\doi{10.1093/biomet/asm023}
\accessdate{Advance Access publication on 4 February 2013}

\received{January 2013}
\revised{April 2013}

\markboth{Zachary T. Kurtz}{Local Log-linear Models for Capture-Recapture}

\title{Local Log-linear Models for Capture-Recapture }

\author{Zachary T. Kurtz}
\affil{Department of Statistics, Carnegie Mellon University, Pittsburgh, Pennsylvania 15213, U.S.A.
\email{zkurtz@stat.cmu.edu}}

\maketitle
\begin{abstract}
Log-linear models are often used to estimate the size of a closed population using capture-recapture data.  When capture probabilities are related to auxiliary covariates, one may select a separate model based on each of several post-strata.  We extend post-stratification to its logical extreme by selecting a local log-linear model for each observed unit, while smoothing to achieve stability.  Our local models serve a dual purpose:  In addition to estimating the size of the population, we estimate the rate of missingness as a function of covariates.  A simulation demonstrates the superiority of our method when the generating model varies over the covariate space.  Data from the Breeding Bird Survey is used to illustrate the method.
\end{abstract}



\section{Introduction}

\citet{Fienberg1972} and \citet{Sanathanan1972} introduced log-linear models for capture-recapture data from closed populations.  Today, log-linear models are often used in applications with 3-6 lists, or capture events to estimate the size of a population of interest.  Recent examples of populations studied include opiate users in Ireland \citep{Kelly2009}, people inside the World Trade Towers on the morning of September 11, 2001 \citep{Murphy2009}, and cases of multiple sclerosis in the Lorraine region of France \citep{Adssi2012}.

Heterogeneity of capture probabilities can cause bias in log-linear models \citep{Darroch1993, Fienberg1999}.  One way to reduce heterogeneity bias is to post-stratify on auxiliary covariates.  In a human population, individuals may be grouped by age, such as 0-10 years, 11-20 years, etc.  A model is fitted to each age group, and the resulting estimates are summed across groups to estimate the population size.  For instance, more than 400 post-strata were used in the evaluation of the coverage of the United States 2000 Census \citep{The2000Census}.

A fundamental challenge in post-stratification is to determining the optimal number of strata.  To remove within-stratum heterogeneity, it is desirable to have as many strata as possible.  At the same time, a minimum stratum size must be maintained to control the variance of estimates.  We partially sidestep this issue by applying log-linear models to individuals, the smallest possible strata, while ``borrowing strength" to maintain adequate effective sample size. 

Specifically, we select and fit a local log-linear model for the capture pattern associated with each individual.  To illustrate this with a human population, suppose that exactly one person of age 19 is observed.  This person constitutes a post-strata of size 1.  A single observation is, of course, not enough to select a log-linear model, but we can get an adequate effective sample size by using a locally weighted average of the capture patterns of people with ages close to 19.  Thus, each local log-linear model is fitted to a local average.

Our approach is closely related to several existing methods.  \citet{Huggins1989} and \citet{Alho1990} developed logistic regression models that allow capture probabilities to vary with auxiliary covariates when there are only two lists.  \citet{Yip2001} extended their method to include certain list interactions when there are more than two lists.  \citet{Chen2002} used ``local post-stratification," which essentially replaces the Alho-Huggins logistic regression with a nonparametric regression.  \citet{Zwane2004} fit log-linear models that used penalized splines to express dependence on a continuous covariate.  With less of an emphasis on interactions between lists, \citet{Hwang2011} used a semi-parametric logistic regression involving local polynomials to model the effect of a continuous covariate, and \citet{Stoklosa2012b} took a similar approach with penalized splines instead of local polynomials.  Our treatment differs from those above by allowing the form -- and not only the fitted values -- of the model to vary over the covariate space.  This generality can meaningfully improve estimates, as we demonstrate via simulation.

Many models treat heterogeneity as a latent feature, without using covariates \citep{Darroch1993, Manrique2008, Pledger2008}.  Such models are especially important when the auxiliary covariates are noninformative or unavailable.  However, when informative covariates are available, their inclusion adds a significant dimension to the value of a capture-recapture study by enabling estimation of the rate of missingness, or the number of unobserved units divided by the number of observed units, at each point in the covariate space.  Thus we learn about not only the population size, but also its composition.  Many existing methods relate covariates to the detection probabilities, but our approach is exceptionally specialized to this task, since we build a full log-linear model at each observed covariate vector.

\section{Notation and general framework}\label{sect:framework}
Suppose $k$ lists, or samples, are drawn from a population of unknown size $n$.  Let $i = 1, ..., n_c$ index the units that are on at least one list.  For each unit $i$ and list $j$, let $y_{ij}$ be the indicator that the $i$th population unit appears on the $j$th list.  Then ${y_{i \cdot} = (y_{i1}, ..., y_{ik})}$ , and $y_{\cdot \cdot}$ is the $n \times k$ matrix with $i$th row $y_{i \cdot}$.  The vector $y_{i \cdot}$ is called the capture pattern of the $i$th unit. Let $x_{i \cdot} $ denote a $1 \times q$ vector of covariates associated with the $i$th unit, and $x_{\cdot \cdot}$ is the $n \times q$ matrix with $i$th row $x_{i \cdot} $.  For each $i > n_c$, the vector $x_{i \cdot}$ is not observed.  If $x_{\cdot \cdot}^c$ is the matrix formed by the first $n_c$ rows of $x_{\cdot \cdot}$, and $y_{\cdot \cdot}^c$ is the matrix formed by the first $n_c$ rows of $y_{\cdot \cdot}$, then the observable data is the pair of matrices $(x_{\cdot \cdot}^c, y_{\cdot \cdot}^c)$.  

Let $\mathcal Y$ denote the set of binary row vectors of length $k$, so each $y_{i \cdot}$  is an element of $\mathcal Y$.  Let $c_{y} := \left| \{i : y_{i \cdot} = y \} \right|$.  The array $c := \{ c_{y}: y \in \mathcal Y \}$ is the contingency table of counts of units in the lists.  In particular, let $c_0 := c_{\vec 0} = n-n_c$, the unknown number of units that are not observed on any list.  Assume that $y_{i \cdot}$ is a realization of a random vector $Y_{i \cdot}$.  Let $p(i, y) = pr( Y_{i \cdot} = y)$, the probability that unit $i$ has capture pattern $y$.  Then $p(i, y_{i \cdot}) = pr(Y_{i \cdot}= y_{i \cdot})$.   

We make a key assumption that is often left unstated in the literature on auxiliary-covariate models of heterogeneity.  Namely, we assume the existence of a function $r( y|x)$ that is piecewise smooth in $x$ and satisfies $p(i, y_{i \cdot}) = r(y_{i \cdot}|x_{i \cdot})\  (i = 1, ..., n$). This is a strong assumption, requiring that the covariates $x$ fully explain any variation in the capture probabilities.  

Let $\vec 0$ denote the row vector of $k$ zeros.  Define the detection function \mbox{$\psi( x) = 1 -  r( \vec 0|x)$}, which is the probability that a unit with covariates $x$ appears in at least one of the lists.  The Horvitz-Thompson estimator of the population size $n$ is
\begin{eqnarray}
\label{eqn:horvThomp}
\tilde n = \sum_{i = 1}^{n_c} \frac{1}{\psi(x_{i \cdot})}.
\end{eqnarray}
The estimator $\tilde n$ relies on the detection probabilities for only the units that are observed.   It is easy to verify that $E (\tilde n) = n$.  Moreover, $\tilde n$ is consistent and asymptotically normal if $\psi(x_{i \cdot})$ is uniformly bounded away from $0$ and $1$  \citep{Alho1990}.  

In practice, using (\ref{eqn:horvThomp}) requires us to estimate the detection function $\psi$.  Our estimator will join several existing capture-recapture estimators that take this route in essence, including those of \citet{Alho1990}, \citet{Chen2002}, and \citet{Zwane2004}.  We begin by putting (\ref{eqn:horvThomp}) into a different form.  Define a function 
\begin{eqnarray}
\label{eqn:pi.definition}
\pi( y|x) :=  \frac{r( y|x) }{ \sum_{ z \neq \vec 0} r( z|x)}   =  \frac{r(y|x) }{ \psi ( x)} .  
\end{eqnarray}
For each nonzero $y \in \mathcal Y$, $\pi( y|x) $ is the conditional probability that a unit with covariates $x$ has capture pattern $y$, given that the unit is observed on at least one list.  

Plugging an estimate $\hat \pi(\vec 0| x)$ of $\pi(\vec 0| x)$ into (\ref{eqn:pi.definition}) and expanding (\ref{eqn:horvThomp}) in terms of (\ref{eqn:pi.definition}) leads us to an estimator involving the sum of the unit-level estimates:
\begin{eqnarray} \label{eqn:newhorv}
\hat n := n_c + \hat c_0, \ \mbox{where }  \ \hat c_0 := \sum_{i =1}^{n_c} \hat \pi(\vec 0| x_{i \cdot}).
\end{eqnarray}
Thus, our challenge is to derive useful estimates $\hat \pi(\vec 0| x_{i \cdot}) \ (i = 1, ..., n_c)$.

We propose the use of local as well as global measures of model performance.  Specifically, let $n_c(x)$ denote the number of units observed at covariate $x$, let $c_0(x)$ denote the corresponding number of missing units, and let $\hat c_0(x)$ denote the corresponding estimate, computed as $n_c(x) \hat \pi(\vec 0|x)$.  Define the local root-mean-square error as 
\begin{eqnarray} \label{eqn:lrmse}
\sqrt{ E[ \{ \hat c_0(x) - c_0(x) \} ^2 ] }.
\end{eqnarray}
Similarly, we use the root-mean-square error $\sqrt{ E(\hat c_0- c_0) ^2 }$ as a global measure of model performance.  These measures can be estimated only in simulations, as validation data typically does not exist in real applications.

Define an (arbitrary) ordering of the capture patterns $ \mathcal Y$ so that $\mathcal Y_j$ denotes the $j$th nonzero capture pattern $(j = 1, ..., 2^k-1)$.  Let $\Pi$ and $A$ be matrices of dimension $n_c \times (2^k-1)$ with elements $\Pi_{ij} = \pi(\mathcal Y_j | x_{i \cdot} )$ and $A_{ij} = I(y_{i \cdot} = \mathcal Y_j)$.  Thus, the $i$th row $A_i$ of $A$ indicates the multinomial outcome corresponding to the multinomial probabilities given in the $i$th row $\Pi_i$ of $\Pi$.  

The vector average $\sum_{i = 1}^{n_c} A_i/n_c$ contains the empirical relative frequencies of the nonzero capture patterns.  This average, together with $n_c$, contains the same information as $c$, and so is sufficient for a traditional log-linear model, as in \citet{Fienberg1972}.  To include heterogeneity that is associated with covariates, we fit local log-linear models to local averages of the form
\begin{eqnarray} \label{eqn:weights}
\widehat \Pi_i = \sum_{t = 1}^{n_c} w^i_t A_t \ (i = 1, ..., n_c),
\end{eqnarray}
where each $w^i$ is a normalized vector of nonnegative weights of length $n_c$.  For example, any basic kernel smoother or weighted k-nearest-neighbors regression can be expressed as (\ref{eqn:weights}).  Although a local average $\widehat \Pi_i$ can be viewed as an estimator of $\Pi_i$ in its own right (hence, the notational similarity), $\widehat \Pi_i$ is used in our framework only as the ``data" for building a local log-linear model.  The $i$th fitted local model, in turn, implies an estimate $\hat \pi(\vec 0| x_{i \cdot})$ as needed for (\ref{eqn:newhorv}). 

The vectors of weights $w^i (i = 1, ..., n_c)$ may be specified by stating a standard kernel and picking a smoothing bandwidth based on subjective researcher expertise.  Although data-driven methods exist for bandwidth selection in local regression, we caution against the careless use of such methods in our specific context for reasons that are discussed in Section \ref{sect:mod.select}.  We take the weights as known and fixed for the remainder of this paper.

For each local average $\widehat \Pi_i$, we define the effective sample size of the $i$th row as 
 \begin{eqnarray} \label{eqn:df}
\eta_i :=  \frac{  \left( \sum_{t=1}^{n_c} w_t^i \right)^2}{ \sum_{t=1}^{n_c} (w_t^i)^2 } = \frac{ 1}{ \sum_{t=1}^{n_c} (w_t^i)^2 } \ (i = 1, ..., n_c).
\end{eqnarray}
This definition, which may have originated with \citet{kish1965}, will be motivated in Section \ref{sect:mod.select}.

\section{Estimating $\pi(\vec 0 , x)$} \label{sect:localLoglin} 
A local log-linear model $\mathcal M_i$ for each $i$th row $\widehat \Pi_i$ is used to estimate the missing cell $\pi(\vec 0 | x_{i \cdot}) \ (i = 1, ..., n_c)$.  While having as many as $n_c$ models for $n_c$ points looks like overfitting in the extreme, it is important to notice that the models are highly correlated across the covariate space, since $\widehat \Pi_i$ is continuous in $x_{i \cdot}$.  If the difference between $x_{i_1 \cdot}$ and $x_{i_2 \cdot}$ is small, than the difference between the local averages $\widehat \Pi_{i_1}$ and $\widehat \Pi_{i_2}$ is also small, effectively constraining $\mathcal M_{i_1}$ to be similar to $\mathcal M_{i_2}$.  Specifically, $x_{i_1 \cdot} = x_{i_2 \cdot}$ implies that $\mathcal M_{i_1} = \mathcal M_{i_2}$.

\subsection{Short review of log-linear models}
Log-linear models provide flexible ways to estimate the multinomial probabilities that underly the cross-classification $c$.  Given a vector of parameters $u = (u_0, u_1, u_2, u_3, u_{12}, u_{23}, u_{13})$, a log-linear model for $c$ when there are $k=3$ lists is
\begin{eqnarray} \label{eqn:satloglinmodel}
\log p( y| u) &=& u_0 + u_1 y_1 + u_2  y_2 + u_3  y_3 +  u_{12} y_1 y_2+  u_{23} y_2 y_3+  u_{13} y_1 y_3,
\end{eqnarray}
where $y_j$ denotes the $j$th element of the vector $y$ $(j = 1,..., k; y \in \mathcal Y_k)$.  The relative probability of being capture on the $j$th list is governed by $u_j$, and the $u_{12}$ term (for example) represents the interaction between the first and second list.  The intercept $u_0$ is of primary interest; the constant $1-\exp(u_0)$ is the probability of detection for an arbitrary unit, and this can be plugged into the denominator of (\ref{eqn:horvThomp}) to get a trivial Horvitz-Thompson style estimator of $n$.

One typically estimates the parameters by maximizing the multinomial conditional likelihood
\begin{eqnarray} \label{eqn:trad.likelihood}
L_c( u | c \setminus c_{ 0}) =  \frac{n_c!}{\prod_{y \neq \vec 0} c_{y}!} \prod_{ y \neq \vec 0} \pi(y| u)^{c_y },
\end{eqnarray} 
where $\pi(y| u) := p(y| u)/(1-p(\vec 0| u))$ \citep{Sanathanan1972}.  

The cross-classification $c$ has only $2^k-1$ observable cells, and a unique maximizer of (\ref{eqn:trad.likelihood}) exists for a model with at most $2^k-1$ parameters.  Thus, if $k=3$, the model (\ref{eqn:satloglinmodel}) is called saturated, and model selection may involve setting some elements of $u$ to be identically zero.   If there are more than three lists, additional parameters may describe higher-order interactions.  

\subsection{Local log-linear models, a special case} \label{sect:locallogfitting}

Fix $i$ in $1, ..., n_c$.  Let $\mathcal W_i$ denote the set of indices of all nonzero entries of the vector of weights $w^i$, and let $n_i = | \mathcal W_i|$, the number of nonzero entries.  We describe a local log-linear model $\mathcal M_i$ for the smoothed data $\widehat \Pi_i$.  Throughout this section we consider only the special case that is defined by the following two assumptions:

\begin{itemize}
\item Boxcar assumption:  $w^i_t = 1/n_i$ for all $t \in \mathcal W_i$.  
\item Local homogeneity:  $r(y|x_{t_1 \cdot}) = r(y|x_{t_2 \cdot})$ for all indices $t_1, t_2 \in \mathcal W_i$ and $y \in \mathcal Y$.
\end{itemize}

The boxcar assumption says simply that the nonzero weights are uniform, which holds for any boxcar kernel smoother or k-nearest-neighbors regression with uniform weights.   The local homogeneity assumption requires that the capture probabilities are constant over $\mathcal W_i$.  Homogeneity is a standard assumption (at least formally) for classical log-linear models.  

By the definition (\ref{eqn:df}), the boxcar assumption gives $\eta_i = n_i$, and the vector
\begin{eqnarray} \label{eqn:boxmulti}
\eta_i \widehat \Pi_i
   &=& \eta_i \sum_{t \in \mathcal W_i} \frac{1}{n_i} A_t 
   = \sum_{t \in \mathcal W_i} A_t
\end{eqnarray}
is a sum of multinomials.  Local homogeneity implies that the $\eta_i$ terms in the sum are identically distributed.  We have already assumed independence between units, and it follows that $\eta_i \widehat \Pi_i$ is a multinomial random variable with $\eta_i$ trials and probabilities $\Pi_i$.  With $k=3$, we apply the saturated local log-linear model (\ref{eqn:satloglinmodel}) to $\eta_i \hat \Pi_i$, replacing the parameter vector $u$ in (\ref{eqn:satloglinmodel}) with a local parameter vector $u_i$.  Since the entries of $\eta_i \widehat \Pi_i$ are the elements of the set $\{ \eta_i \hat \pi(y| x_{i \cdot} ) \}_{y \neq \vec 0}$, the local likelihood function is
\begin{eqnarray} \label{eqn:locLik}
L_i( u^i| \eta_i \widehat \Pi_i) =  \frac{\eta_i!}{\prod_{y \neq \vec 0} \{\eta_i \hat \pi(y| x_{i \cdot}) \}!} \prod_{ y \neq \vec 0} \pi(y| u^i)^{\eta_i \hat \pi(y|x_{i \cdot}) }.
\end{eqnarray} 
Let $\hat u^i$ denote the parameter estimates found by maximizing (\ref{eqn:locLik}).  An important special case is when the kernel is infinite, or when all the units have equal weight so that the local average (\ref{eqn:weights}) coincides with the global average.  Then $\eta_i = n_i = n_c$, and the local likelihood (\ref{eqn:locLik}) coincides with the global likelihood (\ref{eqn:trad.likelihood}).

Various submodels of the saturated model (\ref{eqn:satloglinmodel}) are obtained by removing terms.  The independence model for three lists encodes the assumption that the probability of capture on each list is independent of the event of capture on any other list:
\begin{eqnarray} \label{eqn:indeploglin}
\log \pi ( y| u^i) =& u_0^i + u_1^i  y_1 + u_2^i  y_2 + u_3^i  y_3.
\end{eqnarray} 
We emphasize that model selection is done locally.  In particular, if $i_1 \neq i_2$, the models $\mathcal M_{i_1} $ and $\mathcal M_{i_2} $ may be of completely different forms.  For example, the parameter vector $u^{i_1}$ need not be of the same dimension as $u^{i_2}$.

Given estimated log-linear parameters $\hat u^i$, we estimate $\pi(\vec 0| x_{i \cdot})$ by projecting the corresponding log-linear model onto the missing cell,
\begin{eqnarray} \label{eqn:loglnproj}
\hat \pi(\vec 0| x_{i \cdot}) := \pi ( \vec 0 | \hat u^i) = \exp(\hat u_0^i),
\end{eqnarray} 
and this is all that is needed to construct the population size estimate (\ref{eqn:newhorv}).

\subsection{Local log-linear models, the general case} \label{sect:general}

The likelihood (\ref{eqn:locLik}) was derived from the boxcar and local homogeneity assumptions.  Removing either of these assumptions makes $\eta_i \widehat \Pi_i$ a nontrivial mixture of multinomials, such that (\ref{eqn:locLik}) need not be exactly equal to the probability of the data $\eta_i \widehat \Pi_i$ given the parameters $u^i$.  Relaxing the boxcar assumption means that the local effective sample size $\eta_i$ does not equal the number of nonzero weights $n_i$.  If $\eta_i$ is not integer-valued, exact evaluation of (\ref{eqn:locLik}) requires a continuous generalization of the factorial function, the Gamma function.   Relaxing local homogeneity means that some local heterogeneity in capture probabilities may occur.  Then $\widehat \Pi_i$ is a mixture of not-identically distributed multinomials.  Provided that the true regression function $r(y|x)$ is sufficiently smooth in the covariates $x$, and provided that the bandwidth of the smoother is sufficiently narrow, the corrupting effects of local heterogeneity may be limited.  

Conditional on having selected a model $\mathcal M_i$, maximizing $L_i$ is equivalent to maximizing 
\begin{eqnarray*} \label{eqn:wghtlike}
\sum_{ y \neq \vec 0} \hat \pi(y|x_{i \cdot}) \log \pi(y| u^i)
 &=& \sum_j \hat \pi(\mathcal Y_j|x_{i \cdot}) \log \pi(\mathcal Y_j | u^i) \\
 &=& \sum_{t=1}^{n_c}  \sum_j w_t^i  \log  \{\pi(\mathcal Y_j | u^i)^{I(y_{t \cdot} = \mathcal Y_j)} \},
\end{eqnarray*}
where the final term is a weighted likelihood function, much like the standard weighted likelihood functions that are used for local polynomial regression \citep{loader1999local}.  That is, the approximations involved in relaxing the boxcar and local homogeneity assumptions to use $L_i$ for parameter estimation are analogous to the approximations involved in the use of standard of weighted likelihoods.

\subsection{Local Model Selection} \label{sect:mod.select}
For each $i$th unit, we aim to select the best local model $\mathcal M_i$ within a set of feasible models.  As our set of feasible models, we take all hierarchical local log-linear models that include all main effects.  Let $K_i$ denote the number of free parameters in the $i$th local log-linear model.  For example, the independence model (\ref{eqn:indeploglin}) has four parameters, but $K_i$ is only $3$, since the multinomial probabilities are constrained to sum to $1$.  Define
\begin{eqnarray} \label{eqn:aicc}
IC_i := -2 \log L_i(\hat u^i | \eta_i \widehat \Pi_i) + 2K_i + \frac{2K_i(K_i+1)}{\eta_i -K_i - 1}.
\end{eqnarray}
We interpret $IC_i$ as a local information criterion;  the local log-linear model that gives the lowest value of $IC_i$ is preferred.   Under the boxcar assumption, $IC_i$ is the Akaike information criterion with a small sample bias adjustment by \citet{hurvich1989regression}, which has performed well in capture-recapture settings \citep{burnham1995model, anderson1999understanding}. 

Without the boxcar assumption, our criterion (\ref{eqn:aicc}) serves only as an approximate generalization of the corrected Akaike information criterion.  In fact, the definition of the effective local sample size (\ref{eqn:df}) is motivated to make $\eta_i \widehat \Pi_i$ as similar as possible to a multinomial distribution so that inference using (\ref{eqn:aicc}) is sensible.  In particular, $\eta_i$ has the property that the element-wise (i.e., binomial) variances of $\eta_i \widehat \Pi_i$ are the same as the element-wise variances of an exactly multinomial random variable with $\eta_i$ trials and outcome probabilities  $\Pi_i$.  

The choice of kernel bandwidth for the weights in the local average $\widehat \Pi_i$ is intimately related to the selection of local log-linear models.  If the bandwidth for $\widehat \Pi_i$ is too small, then $\eta_i$ is small, and the criterion $IC_i$ tends to favor an extremely sparse local model even if important high-order interactions are present.  When the bandwidth is large, the criterion $IC_i$ tends to favor a local model with many parameters.  Since increasing the bandwidth can correspond to increasing model complexity, the usual bias-variance tradeoff that traditionally guides bandwidth selection for kernel regression is reversed through the local model selection step.   Thus, in addition to bias and variance, one must consider the tradeoff between a bandwith that is small enough to reduce heterogeneity and yet large enough to facilitate the selection of local log-linear models with enough parameters to capture dependencies between lists.  This intuition guides our ad-hoc choices of bandwidth, and we leave data-driven bandwidth selection methods for future work.

One aspect of our criterion (\ref{eqn:aicc}) bears sharp contrast with previous uses of information criteria for problems involving data smoothing.  In a typical nonparametric regression problem, the sample size is unambiguous, and the number of parameters in the model is approximated as the trace of the ``hat" matrix.  Our emphasis on local model selection leads to the converse:  the number of parameters can be counted directly, but the sample size must be approximated as $\eta_i$, which is based on the $i$th row of the ``hat" matrix.

\section{Bootstrap variance estimation}

Treating $\eta_i \widehat \Pi_i$ as an approximately multinomial random variable, the asymptotic variance formulas of \citet{Fienberg1972} could be relevant for specific local log-linear models.  However, \citet{Norris1996b} emphasized the importance of including model uncertainty in the variance estimate.  \citet{Norris1996b} proposed several bootstrap methods that include model uncertainty in capture-recapture settings; we adopt their ``Method 2" to estimate the unconditional variance of the population size estimate $\hat c_0$ defined in (\ref{eqn:newhorv}).   We describe the method in detail, including modifications that are needed to deal with the auxiliary covariates.  

The first step is to simulate covariate vectors of the unobserved units.  Together with the observed covariates $x_{\cdot \cdot}^c$, these new covariate vectors define a population that is consistent with the model.  The second step is to randomly assign a capture pattern for each unit, discarding all units that are assigned the $\vec 0$ capture pattern.  The third step is to select and fit local log-linear models to the simulated data to obtain a bootstrap estimate $\hat c_0^{boot}$.  Replicating the bootstrap $B$ times gives a set $\{\hat c_0^{boot}(1), ..., \hat c_0^{boot}(B) \}$, and the variance of this set is the bootstrap estimate of $Var(\hat c_0)$.  The following subsections provide details on the first two steps.

\subsection{Simulating unobserved units for the bootstrap}
The covariate matrix $x_{\cdot \cdot}^c$ contains a row for the covariate vector of each observed unit.  We simulate approximately $\hat c_0$ additional covariate vectors to represent the unobserved units.  According the fitted local model, $\hat \pi(\vec 0 | x_{i \cdot})$ is the number of unobserved units with covariate vector $x_{i \cdot}$.  

However, $\hat \pi(\vec 0 | x_{i \cdot})$ is not generally an integer, and it is not clear how to interpret non-integer numbers of units.  Much like \citet{Zwane2003}, we use random rounding to replace $\hat \pi(\vec 0 | x_{i \cdot})$ with a whole number, as follows.  Decompose each $\hat \pi(\vec 0 | x_{i \cdot})$ into its integer and decimal parts, $\hat \pi_i^{int}$ and $\hat \pi_i^{dec}$, such that \mbox{$\hat \pi(\vec 0 | x_{i \cdot})  = \hat \pi_i^{int}+ \hat \pi_i^{dec} $}.  Let $\tilde \pi_i$ denote the result of randomly rounding $\hat \pi(\vec 0 | x_{i \cdot})$, where one rounds up to the next larger integer $\hat \pi_i^{int}+1$ with probability $\hat \pi_i^{dec}$, and rounds down to the next smaller integer $\pi_i^{int}$ with probability $1- \hat \pi_i^{dec}$.  

Let $c_0^{sim} = \sum_i \tilde \pi_i$, noting that $E(c_0^{sim}) = \hat c_0$.  Let $x_{\cdot \cdot}^{sim}$ denote a $c_0^{sim}  \times q$ matrix of covariate row vectors that are copied from $x_{\cdot \cdot}^c$ according to the nonzero elements of $\{\tilde \pi_i : i = 1, ..., n_c \}$.  For example, if $\tilde \pi_7 = 2$, then $x_{\cdot \cdot}^{sim}$ contains two rows which are replicates of the $7$th observed covariate vector $x_{7 \cdot}$. Finally, let $x_{\cdot \cdot}^{c+sim}$ denote the $(n_c + c_0^{sim})  \times q$ matrix of covariates formed by appending $x_{\cdot \cdot}^{sim}$ to the bottom of $x_{\cdot \cdot}^c$.  The matrix $x_{\cdot \cdot}^{c+sim}$ represents the full population to be used for the bootstrap.
The number of new units $c_0^{sim}$ tends to be close, but not generally equal, to $\hat c_0$.  Thus, random rounding introduces some variability that is not part of the modeling process, and this may slightly inflate the bootstrap variance, leading to conservative confidence intervals.

\subsection{Assigning capture patterns for the bootstrap}

Building on the definition (\ref{eqn:pi.definition}), define estimates for $r(y|x_{i \cdot})$ as
$$\hat r(y|x_{i \cdot}) := \hat \pi(y|x_{i \cdot}) \hat \psi(x_{i \cdot}) = \frac{ \hat \pi(y|x_{i \cdot})}{1+ \hat \pi(\vec 0|x_{i \cdot}) } \ (i = 1, ..., n_c+c_0^{sim};  y \in \mathcal Y),$$
where $\hat \pi(y|x_{i \cdot})$ is an element of $\widehat \Pi_i$ if $y \neq \vec 0$, and $\hat \pi(\vec 0|x_{i \cdot})$ is defined in (\ref{eqn:loglnproj}).  Draw the capture pattern for the $i$th unit from the set $\mathcal Y$ of possible capture patterns according to the multinomial probabilities $\{ \hat r(y|x_{i \cdot}) : y \in \mathcal Y\}$. 

\section{Bird species richness}  
As a simple illustration, we estimate the number of bird species in continental North America, north of Mexico, using three years of data from the North American Breeding Bird Survey  \citep{Sour2011}.  Table \ref{tab:birds} displays the cross-classification of species observed in the years 2009 - 2011, treating each year as a separate list.  For instance, the table shows that 581 species were observed in all three years, and 18 species were observed only in 2009.  

Define a covariate $x$ as the reverse of the rank ordering among the $664$ observed species based on the total number of times that each species was observed over the three year period.  For example, the species that was observed most often has the largest possible value, $x = 664$.  The obvious interpretation of $x$ is that species with a high value of $x$ are easy to observe.  Compared to covariates used previously to model bird populations, such as wingspan, our covariate appears to be a relatively direct proxy for species detectability.

\begin{table}[h!]  
\caption{Cross-classification of species by years observed }
\centering
    \begin{tabular}{crcc}
     
                                          ~ &         ~ &  In 2011   &  Not in 2011  \\ 
       \multirow{2}{*}{In 2010} & In 2009 & $581$ & $13$  \\
                                                 & Not in 2009 & $10$ & $10$  \\
      \multirow{2}{*}{Not in 2010}& In 2009 & $11$ & $18$  \\
                                               & Not in 2009 & $21$ & $c_0$  
    \end{tabular} 
 \label{tab:birds}
\end{table}

\begin{figure}[h!]
\centering
\includegraphics[width =  \textwidth]{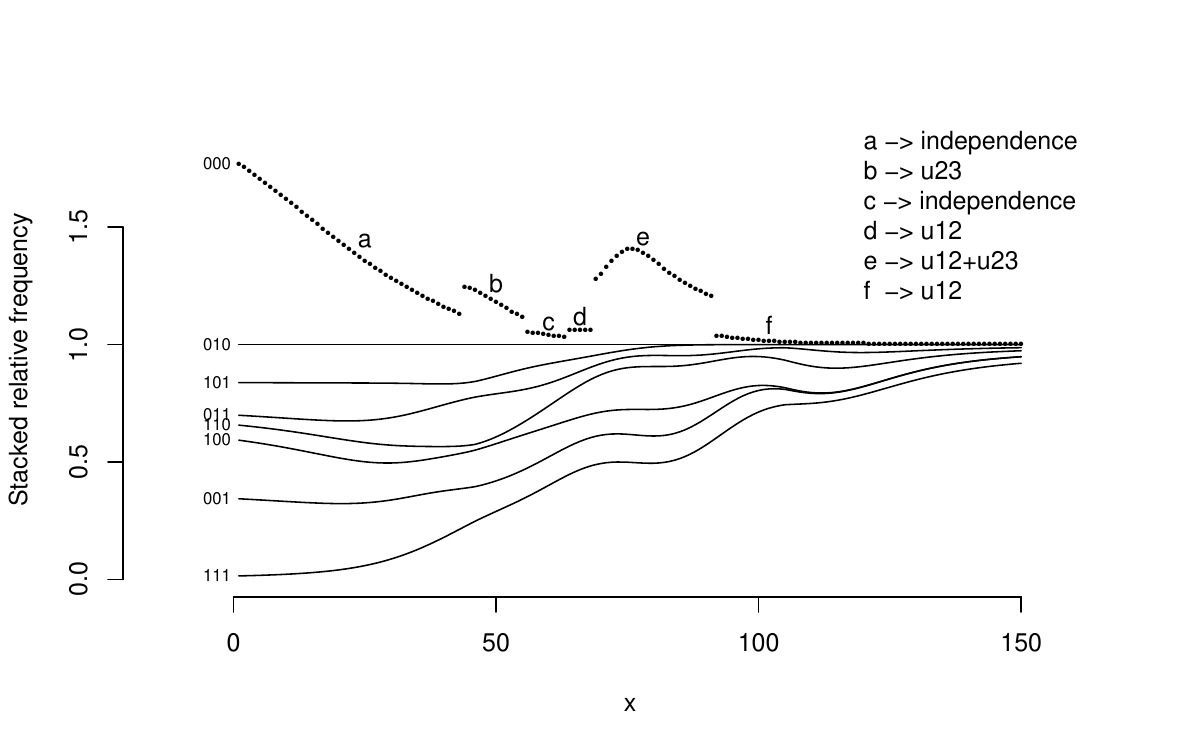}
\caption{Our proxy for species detectibility, $x$, is on the horizontal axis.  The relative frequencies of the capture patterns (i.e., ``111", ``001", ...) are plotted as functions of $x$ in a stacked form.  For example, the curve labeled ``001" represents the sum \mbox{$ \pi \{ 111| x \} +  \pi \{ 001| x \}$}.  The relative frequencies of observable capture patterns sum to 1, the horizontal line, labeled ``010".  Above this horizontal line, the estimates $\hat \pi( \vec 0| x_i)$ are plotted as $\hat \pi( \vec 0| x_i) + 1 \ (i = 1, ..., 150)$.  Each continuous section of that curve corresponds to a specific log-linear model, as indicated by the labels {\em a - f}.  The ``independence" model is (\ref{eqn:indeploglin}), and ``u23" is shorthand for the log-linear model that includes the interaction coefficient $u_{23}$ in addition to the main effects. }  
\label{fig:birds}
\end{figure}

We estimate the $\pi(\vec 0|x_i)$ only for $i = 1, ..., 150$, corresponding to the $150$ least-observed species, since the species that are difficult to observe are the only ones for which significant numbers of species can have gone missing.  Weights from a Gaussian kernel define the local averages (\ref{eqn:weights}).  The bandwidth is set at $45$ but is increased near the boundary such that the number of nonzero weights is constant across units.   The local averages $\{ \hat \pi(y | x) \}_{y \neq 0}$ are plotted in a stacked form as seven smooth functions of $x$ in Figure \ref{fig:birds}. 

The information criterion (\ref{eqn:aicc}) is minimized to select a local log-linear model for each observed species.   The resulting log-linear estimates for $\pi(\vec 0| x)$ are plotted as individual points above the horizontal line at $1$ in Figure \ref{fig:birds}.  These estimates appear to follow a curve that has discontinuities at the points at which a different model is selected.

We interpret each estimate $\hat \pi(\vec 0| x_i)$ as a rate of missingness.  For example, at $x = 1$, corresponding to the least frequently observed species, the distance between the uppermost point (labelled ``000") and the horizontal line below it is approximately 0$\cdot$7, indicating an estimated rate of missingness $\hat \pi(\vec 0|x=1) \approx 0.7$.   For all units with $x \geq 100$, the rate of missingness is nearly zero.

Applying (\ref{eqn:newhorv}) gives $\hat c_0 = 28.8$, or $\hat n = 692.8$.  The bootstrapped 95\% confidence interval for $\hat c_0$, based on 500 replications is $(12.4, 30986)$.  The large right tail of the confidence interval is possibly related a point made by \citet{Alho1990}:  the Horvitz-Thompson sum is unstable when the detection probability $\psi( x)$ approaches $0$, even if $\psi( x)$ were known.  Indeed, the detection probabilities in the left tail of the distribution of $x$ in Figure \ref{fig:birds} may be too low to estimate accurately.  This point deserves more attention in many capture-recapture studies, including previous studies using Breeding Bird Survey data such as \citet{Boulinier1998} and \citet{Dorazio2003}.

The Breeding Bird Survey data exists for many years prior to 2009, and so our use of only three years of data raises an obvious question:  Why not extend the model to incorporate all available years?  However, the assumption of a closed population may fail over long spans of time, as certain species go extinct, and new species evolve or change their geographic region of preference.  Population size estimation on a 3-year moving window could, in principle, reveal changes in species richness over time.  A separate consideration is that not using data earlier than 2009 allows us to use the previous years of data as a partial validation of our method.  The data from 1997 to 2008 includes 40 additional species, which is on a similar order of magnitude as our point estimate.

\section{A Simulation Experiment}  
\begin{figure}[h!]
\centering
\includegraphics[width = \textwidth]{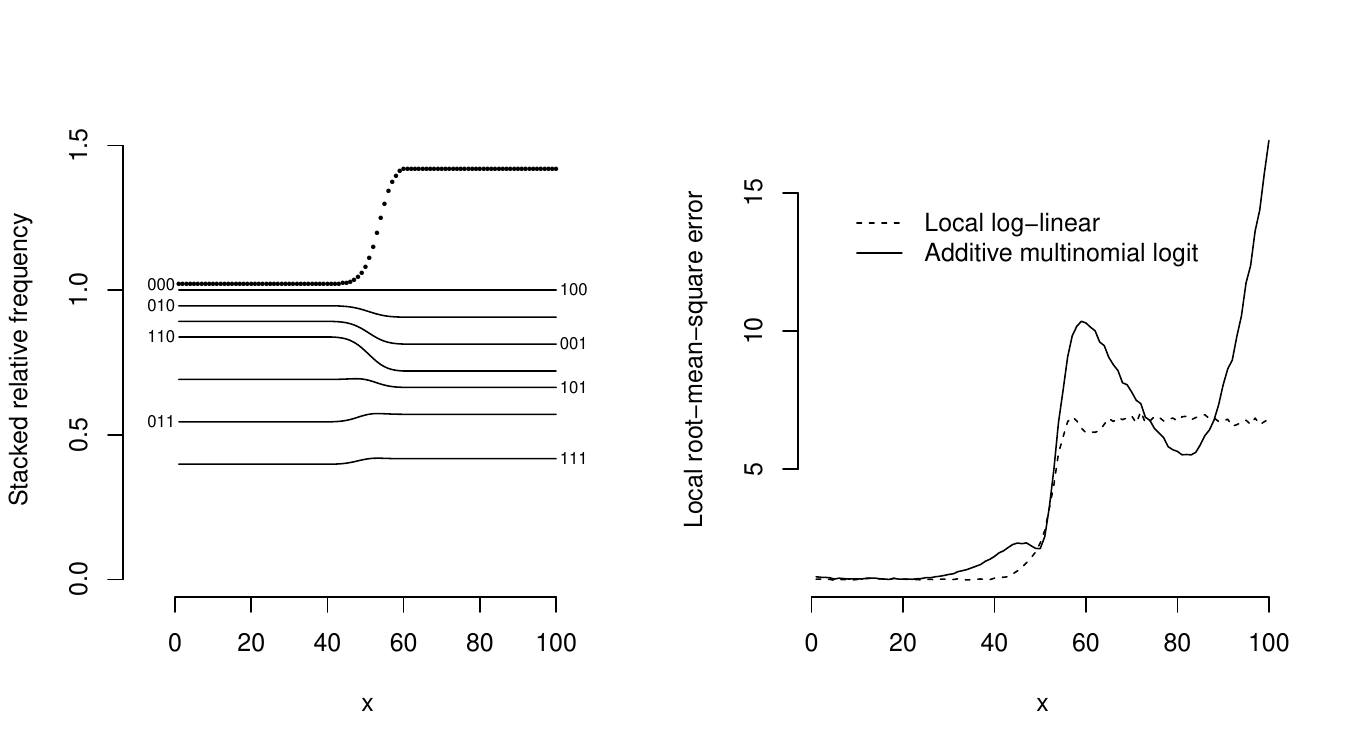}
\caption{The first panel illustrates the probability structure of the generating model.  See the caption of Figure \ref{fig:birds} for the basic interpretation of the stacked curves.  The dotted curve, labeled ``000" indicates the rate of missingness at each value of $x$ for the generating model.  For example, when $x<40$, the rate of missingness is less than $0.05$.  The second panel shows the local root-mean-square error (\ref{eqn:lrmse}) for local log-linear models and for the additive multinomial logit model.  }  
\label{fig:simProbs}
\end{figure}

We conducted a simulation experiment to compare local log-linear models against the additive multinomial logit model \citep{Zwane2004} when the form of the generating model varies over a single continuous covariate.   In each simulation, a population of size $\mbox{n = 5000}$ is spread uniformly over the integers $x=1, ..., 100$.  The fact that many (i.e, $50$) units are assigned to each unique covariate lightens the computational burden for local log-linear models, since it suffices to select a single local model at each of the $100$ possible values of $x$ instead of selecting a model for each of several thousand observed units. The generating model for simulating the capture patterns is defined  in terms of two log-linear models for three lists,
\begin{eqnarray*}
\log p_v(y) &=&  v_0 + v_1 y_1 + v_2 y_2 + v_3 y_3 +  v_{12} y_1 y_2  + v_{13} y_1 y_3  + v_{23} y_2 y_3\\
\log p_w(y) &=&  w_0 + w_1 y_1 + w_2 y_2 + w_3 y_3 +  w_{12} y_1 y_2  + w_{13} y_1 y_3  + w_{23} y_2 y_3.
\end{eqnarray*}
Let 
\begin{eqnarray*}
v &=& (v_1, v_2, v_3, v_{12}, v_{13}, v_{23}) = (1, 1, 1, 0, 0, 0)\\
w &=& (w_1, w_2, w_3,w_{12}, w_{13}, w_{23}) = (-1.5, -1.5, 1, 1.5, 2)
\end{eqnarray*}
Note that the first log-linear model has no list interactions, and the second model has interactions between every pair of lists.  The generating model for the simulation is defined in terms of $v$ and $w$, beginning with model $v$ and transitioning smoothly to model $w$, as displayed in the first panel of Figure \ref{fig:simProbs}.  The figure shows the relative frequencies $\{ \pi(y|x) \}_{y \in \mathcal Y}$, defined in terms of $p(y|x)$ as in (\ref{eqn:pi.definition}), in a stacked form.  

To obtain the smooth transition shown in the first panel of Figure \ref{fig:simProbs}, let $\Phi$ denote the cumulative distribution function of a standard normal random variable, and (somewhat arbitrarily) let
$$T(x) = \frac{ \Phi \{5(x - 10)/20 \} - \Phi (-5*9/20) }{  \Phi (2.5) -  \Phi (-2.25)}  \ (x = 1, ..., 20).$$
For the generating model we take $u(x) = \{ u_1(x), u_2(x), u_3(x), u_{12}(x), u_{13}(x), u_{23}(x) \}$ as
$$u(x) = 
\left\{
	\begin{array}{ll}
		v & \ (x = 1, ..., 40) \\
		 \{1-T(x) \}v  +  T(x) w  & \ (x = 41, ..., 60) \\
		w & \ (x = 61, ..., 100).
	\end{array}
\right.
$$
The multinomial capture probabilities are defined at each $x$ according to the log-linear model (\ref{eqn:satloglinmodel}), and $u_0(x)$ is uniquely determined by the requirement that the multinomial probabilities must sum to $1$.  

We simulated $2000$ replications of the capture-recapture experiment, with the capture patterns drawn as independent multinomials according to the relative capture pattern frequencies illustrated in the first panel of Figure \ref{fig:simProbs}.   On each replication of the experiment, we performed local log-linear modeling, with weights defined using the Epanechnikov kernel with a bandwidth of $12$, producing local effective sample sizes on the order of $600$.  The exact choice of bandwidth was arbitrary, but the order of magnitude was motivated by relatively basic (i.e., without covariates) simulation experiments that seemed to suggest that sample sizes of several hundred are needed to produce reasonably stable model selection results for log-linear models with three lists.  In a post-hoc analysis, we repeated the simulations with the kernel bandwidth set to $10$ and $14$, resulting in slightly worse and significantly better performance, respectively. 

For comparison, we replicated the implementation of the additive multinomial logit model as in \citet{Zwane2004}, who used the \texttt{VGAM} package \citep{yee2010vgam} in \texttt{R}.  In addition, we partitioned the data into post-strata of approximately equal size and selected a log-linear model on each stratum using the Akaike information criterion with the small sample correction of \citet{hurvich1989regression}.  Table \ref{tab:simres} summarizes the performance of each model for  the task of estimating $c_0$, the number $c_0$ of units that were not captured in each simulation. The post-stratification with 5 post-strata had the best results among several numbers of strata that we tried.   Table \ref{tab:simres} shows all three methods performing comparably well.  To put the biases into perspective, $c_0$ was typically around $1300$ in these experiments. 

\begin{table}[h]  
\caption{Simulation results}
\centering
    \begin{tabular}{rrrrr}
Model & Root MSE & Bias & Standard deviation &   95\% Interval width \\
Local log-linear                      & 149   &    -8       &  149     &    579\\
Additive multinomial logit & 152   &   -62     &  139      &    547\\
Log-linear, 5 post-strata     & 153     &   14     &  152      &    593
    \end{tabular} 
 \label{tab:simres}
\end{table}

The local root-mean-square error (\ref{eqn:lrmse}) is an important aspect of model performance that is not reflected in Table \ref{tab:simres}.  We compute the empirical local root-mean-square error at each $x$ across the $2000$ simulation replications and plot the result in the second panel of Figure \ref{fig:simProbs}.  We conclude that local log-linear models can outperform the additive multinomial logit model in terms of the local error in certain settings.

\section{Discussion}

Local log-linear models point to several avenues of future work.  First, in Section \ref{sect:mod.select} we identified the desirability of a data-driven model selection criterion that simultaneously optimizes the local averaging bandwidth and the complexity of the local models.  Second, the information criterion (\ref{eqn:aicc}) is in need of a more detailed theoretical basis.  In particular, the effective local sample size is defined heuristically and may not be optimal.  Third, variations and refinements to traditional log-linear models have been suggested by  \citet{cormack1989log}, \citet{Darroch1993}, and \citet{rivest2001improved}, and could be integrated with our approach.  Fourth, our discrete method of local log-linear model selection (Section \ref{sect:mod.select}) could be improved upon with local model averaging, or even a fully Bayesian approach as in  \citet{madigan1997bayesian}.  Of course, building a fully Bayesian model separately for each observed unit may require superior computational speed for even a dataset of moderate size.

Our simulation to compare local log-linear models against the additive multinomial logit model should not be interpreted as an endorsement of local models for every application.  Local models may have unique ability to accurately estimate rates of missingness in a large and diverse population, such as the human population of a nation, in which the basic relationships between lists may vary across age and socioeconomic group.  Large sample sizes are obviously required to reliably estimate a large set of unique local models.  However, even when the sample size is large, there may be scientific reasons to believe that a single log-linear relationship should hold across all strata, and in this case the additive multinomial logit is expected to be superior.

\section*{Acknowledgements}
Stephen E. Fienberg provided expert opinion and relevant references.  Cosma Rohilla Shalizi provided generous technical and editorial advice.  William F. Eddy and Rebecca Steorts made countless contributions on style and content.  This work was partially supported by the NSF.

\bibliographystyle{biometrika}
\bibliography{../myrefs}

\end{document}